\documentclass{IEEEtran4PSCC}
\ifCLASSINFOpdf
   \usepackage[pdftex]{graphicx}
\else
   \usepackage[dvips]{graphicx}
\fi
\usepackage[cmex10]{amsmath}
\usepackage[cmex10]{amsmath}

\usepackage{cite}
\usepackage{amsmath,amssymb,amsfonts}
\usepackage{algorithmic}
\usepackage{graphicx}
\usepackage{textcomp}
\usepackage{xcolor}
\usepackage{multicol, blindtext}
\usepackage{xspace,empheq,fancybox,amsmath,amssymb,graphicx,epstopdf,epsfig,syntonly,times,amsthm} \usepackage{psfrag,color,bm,array,cite}
\usepackage{url,cite,footnote,xspace,syntonly,algorithm,algorithmic}
\usepackage{verbatim,multirow}
\usepackage[T1]{fontenc}
\usepackage{mathtools}
\usepackage{amsmath,amsfonts,amsthm,bm}
\usepackage{graphicx}
\usepackage{mwe}
\usepackage{caption}
\usepackage{stfloats}
\usepackage{amsfonts}
\usepackage{stackengine}
\usepackage{lipsum}
\usepackage{subfig}

\def\BibTeX{{\rm B\kern-.05em{\sc i\kern-.025em b}\kern-.08em
    T\kern-.1667em\lower.7ex\hbox{E}\kern-.125emX}}
    
\newtheorem{remark}{\bfseries Remark}

\newcolumntype{P}[1]{>{\centering\arraybackslash}p{#1}}

\definecolor{color_yuqi}{RGB}{39, 155, 228}

\makeatletter
\let\old@ps@headings\ps@headings
\let\old@ps@IEEEtitlepagestyle\ps@IEEEtitlepagestyle
\def\psccfooter#1{%
    \def\ps@headings{%
        \old@ps@headings%
        \def\@oddfoot{\strut\hfill#1\hfill\strut}%
        \def\@evenfoot{\strut\hfill#1\hfill\strut}%
    }%
    \def\ps@IEEEtitlepagestyle{%
        \old@ps@IEEEtitlepagestyle%
        \def\@oddfoot{\strut\hfill#1\hfill\strut}%
        \def\@evenfoot{\strut\hfill#1\hfill\strut}%
    }%
    \ps@headings%
}
\makeatother

\begin{document}
%

\title{A Data-Driven Approach for High-Impedance Fault Localization in Distribution Systems}

\author{
\IEEEauthorblockN{Yuqi Zhou$^{*}$,  Yuqing Dong$^{\dag}$, and Rui Yang$^{*}$}
\IEEEauthorblockA{$^{*}$National Renewable Energy Laboratory, Golden, CO, USA\\
$^{\dag}$The University of Tennessee, Knoxville, TN, USA\\}
}


\maketitle

\begin{abstract}
Accurate and quick identification of high-impedance faults (HIFs) is critical for the reliable operation of distribution systems. Unlike other faults in power grids, HIFs are very difficult to detect by conventional overcurrent relays due to the low fault current. Although HIFs can be affected by various factors, the voltage–current characteristics can substantially imply how the system responds to the disturbance and thus provides opportunities to effectively localize HIFs. In this work, we propose a data-driven approach for the identification of HIF events. To tackle the nonlinearity of the voltage–current trajectory, first, we formulate optimization problems to approximate the trajectory with piecewise functions. Then we collect the function features of all segments as inputs and use the support vector machine approach to efficiently identify HIFs at different locations. Numerical studies on the IEEE 123-node test feeder demonstrate the validity and accuracy of the proposed approach for real-time HIF identification.
\end{abstract}


\thanksto{\protect\rule{0pt}{0mm} 
This work was authored in part by the National Renewable Energy Laboratory, operated by Alliance for Sustainable Energy, LLC, for the U.S. Department of Energy (DOE).
Funding for Y. Zhou and R. Yang provided by U.S. Department of Energy Office of Energy Efficiency and Renewable Energy Solar Energy Technologies Office (SETO) under award number DE-EE0038408.
The views expressed in the article do not necessarily represent the views of the DOE or the U.S. Government. The U.S. Government retains and the publisher, by accepting the article for publication, acknowledges that the U.S. Government retains a nonexclusive, paid-up, irrevocable, worldwide license to publish or reproduce the published form of this work, or allow others to do so, for U.S. Government purposes.
}

\section{Introduction}

High-impedance faults (HIFs) occur when an energized conductor comes into contact with a high-impedance medium (e.g., sod, asphalt, tree limb). As a consequence of the low-fault current therein, HIFs are very difficult to detect by conventional overcurrent protection devices. In addition, if left unattended, such faults could pose fire risks and even cause electric shock, endangering human life. Therefore, the timely localization of HIFs in distribution systems is crucial for safe and reliable grid operation.

Because the impedance of HIFs is highly uncertain and can fluctuate over time, it is very difficult to deterministically model the dynamic process of arcing. Meanwhile, limited monitoring devices in the distribution system make it challenging to pinpoint the HIF locations. Due to these restrictions, machine learning approaches are uniquely positioned for HIF localization.
Early work used neural network algorithms \cite{sirojan2018sustainable,wang2020use} to identify the existence of HIFs in the distribution system and achieve satisfactory performance. More recently, autoencoder-based approaches \cite{rai2021deep,li2021physics} have been explored for HIF learning with partially labeled data sets. Albeit sufficiently accurate, in general, these approaches lack explainability and interpretability on the detection model. In contrast, algorithms like support vector machines (SVMs) are suitable for producing more explainable solutions under varying operating conditions.
Earlier efforts have been made to localize faults using SVM-based approaches \cite{sarwar2020high,ahmed2023hierarchical}, but thus far the dynamic electrical characteristics are not suitably integrated into the model.

To address this issue, we propose an efficient and explainable SVM approach for localizing HIFs in the distribution system. For dimension reduction, we formulate tractable optimization problems to approximate the voltage-current (V-I) trajectory with piecewise functions. Notably, the proposed optimization method can be easily modified to incorporate different approximation requirements. We collect the properties of the piecewise functions (e.g., slope rates in the piecewise linear approximation) and directly use them as the inputs to the SVM model. Meanwhile, the learning outputs are the HIF locations (e.g., bus number). The approximated functions can be reckoned as a unique projection from the original nonlinear trajectory, and the properties of these functions are indicative of the electrical characteristics of HIFs during arc quenching and restriking. Hence, the simplified learning inputs not only nicely conserve the unique properties of HIF arcing but also effectively reduce the dimensionality of the measurements from the original nonlinear trajectory.
Admittedly, even for faults at the same location, their V-I trajectories might be different due to operation conditions and measurement noise. However, the rationale behind our proposed approach is that the electrical characteristics of faults at the same location share similarities, and these are reflected in the properties of the piecewise approximation function and can be effectively captured by the SVM algorithm.

This paper is organized as follows. Section \ref{sec:HIF} introduces the dynamic model of HIFs. Section \ref{sec:approx} presents the optimization formulation to obtain the optimal piecewise approximation. Section \ref{sec:svm} explains the SVM algorithm and how to use approximation results for the HIF identification task. Section \ref{sec:NR} shows numerical simulations on the IEEE 123-node test feeder, and Section \ref{sec:con} concludes the paper.



\section{High-Impedance Fault Model}
\label{sec:HIF}

\begin{figure}[h!]
\centering
\includegraphics[trim=0.1cm 0cm 0cm 0.2cm,clip=true,totalheight=0.16\textheight]{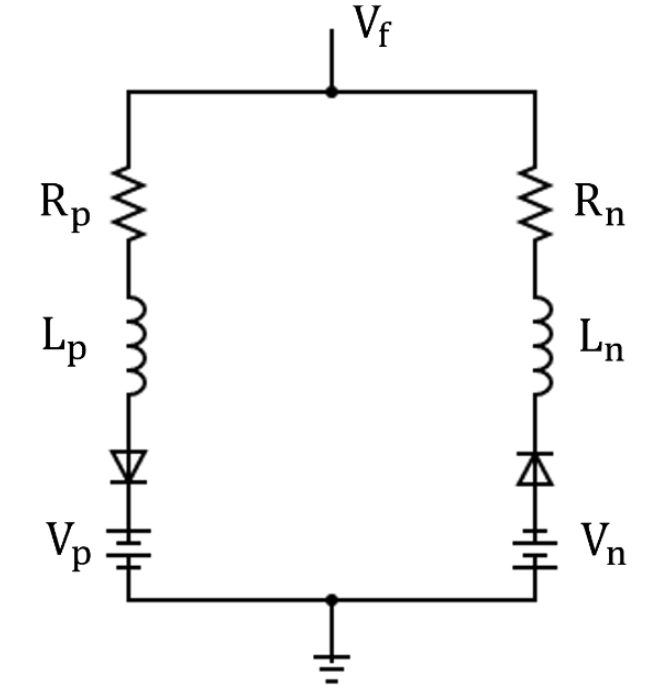}
\caption{HIF modeled with two antiparallel diodes}
\label{fig:HIF_model_new}
\vspace*{-1mm}
\end{figure}

Accurately modeling HIFs is crucial for the efficient identification of these events.
Although various arc models of HIFs have been studied, the model using two antiparallel DC voltage sources and diodes \cite{cui2019feature,gautam2012detection} is well-known to be able to effectively simulate the dynamic activities of arcs during HIF events, as shown in Fig.~\ref{fig:HIF_model_new}.

When $V_{f} > V_{p}$, the current flows from the source to the ground. On the other hand, when $V_{f} < V_{n}$, the negative cycle occurs, during which the current flows back to the source. As $V_{n} < V_{f} < V_{p}$, there is no current in the HIF circuit, which represents the arc quenching.
When the voltage $V_{f}$ falls within these ranges, the electrical characteristics at the fault location as well as the rest of system are generally different. As a matter of fact, the voltage and current at the distribution feederhead approximately feature a piecewise trajectory \cite{wang2016high} during the HIF events. 
The occurrences of HIFs at different system locations will cause the voltage-current (V-I) trajectories at the feederhead to differ as well.
The uniqueness of the voltage-current (V-I) trajectory has been recognized to be very helpful for HIF detection and localization tasks. Nonetheless, the randomness and nonlinearity of the HIF process make it extremely challenging to attain a closed-form expression for the V-I trajectory. This motivates us to explore more tractable and efficient algorithms for the initial approximation of the V-I trajectory before integrating it into the learning framework.

\section{Piecewise Approximation of V-I Trajectory} \label{sec:approx}


Providing tractable yet accurate approximations of the V-I trajectory is crucial for later identifying the fault locations.
In this section, we present optimization formulations for finding the optimal piecewise linear and piecewise quadratic approximations, both of which can be efficiently handled by mathematical solvers. Notably, the problems are shaped to provide continuous approximations over pieces instead of optimizing each one individually, which is more logical and explainable especially for the V-I trajectory approximation.


\begin{remark}[Segmentation for the approximation]
In this paper, we approximate the nonlinear trajectory with three segments as an example, because the V-I characteristics of HIFs roughly feature a piecewise function composed of three pieces (see, e.g., \cite{cui2019feature,wang2016high}). Nonetheless, the formulation can be easily adjusted to incorporate more pieces. 
\end{remark}

\subsection{Piecewise Linear Approximation}

\begin{figure}[t!]
\centering
\includegraphics[trim=0cm 0cm 0cm 0.5cm,clip=true,totalheight=0.16\textheight]{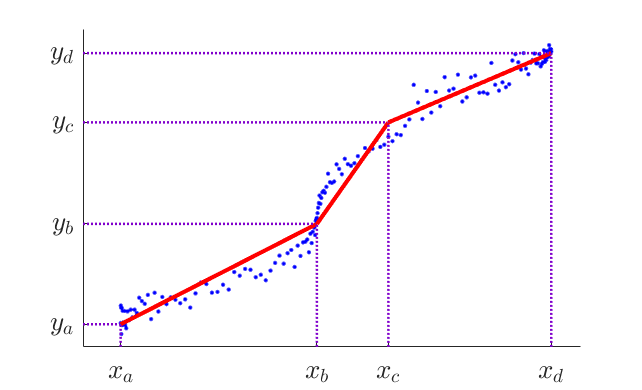}
\caption{Illustration of the piecewise linear approximation}
\label{fig:approximation_linear}
\vspace*{-3mm}
\end{figure}

The piecewise linear approximation is a regression model that aims to approximate different pieces of the trajectory using affine functions (see Fig. \ref{fig:approximation_linear}). Without loss of generality, we assume that the breakpoints $x_a, x_b, x_c, x_d$ for the linear approximation are known parameters. Our goal is to find three affine functions, as follows, that optimally fit the data points $(I_{i}, V_{i})$ from the trajectory:
\begin{align}
y_{k} = s_{k}x + h_{k}, \quad \forall k = 1, 2, 3.
\end{align}
The slope rates for these affine functions are $s_{1} = \frac{y_{b}-y_{a}}{x_{b}-x_{a}}$, $s_{2} = \frac{y_{c}-y_{b}}{x_{c}-x_{b}}$, and $s_{3} = \frac{y_{d}-y_{c}}{x_{d}-x_{c}}$, respectively. Since two points are sufficient to uniquely determine an affine function and the $x$-values for the breakpoints are given, the problem can be simplified to finding the optimal $y_a$, $y_b$, $y_c$, $y_d$ values. To this end, the piecewise linear approximation can be formulated as an optimization problem to minimize the total distance between the data points and the fitting affine functions:
\begin{subequations} \label{eq:LR}
\begin{align} 
\min \quad & \sum^{3}_{k = 1} \sum^{N_k}_{i = 1} \left(s_{k}I_{i} + y^{*}_{k} - s_{k}x^{*}_{k} - V_{i}\right)^{2} \label{eq:LR_a}\\
\textrm{s.t.} \quad 
  & \underline{y}_{\zeta} \leq y_{\zeta}  \leq \overline{y}_{\zeta}, \quad \zeta = a, b, c, d. \label{eq:LR_c}
\end{align}
\end{subequations}
where we use $N_{k}$ to denote the number of data points in the $k$-th segment. The fitting distance for piece $k$ is equal to $\sum^{N_k}_{i = 1}\left(s_{k}I_{i} + h_{k} - V_{i}\right)^{2}$, and substituting $h_{k} = y^{*}_{k} - s_{k}x^{*}_{k}$ yields the expression in objective function \eqref{eq:LR_a}, where $(x^{*}_k, y^{*}_k)$ is the breakpoint in the $k$-th piece function. For instance, when $k = 1$, $(x^{*}_k, y^{*}_k)$ can be either $(x_a, y_a)$ or $(x_b, y_b)$. The limits of decision variables are enforced in constraints \eqref{eq:LR_c}. Note that these constraints are provided in the optimization formulation as a preventive measure for data anomalies, but they are not strictly necessary in general.
Because $I_{i}, V_{i}, x_{k}$ are known constants and slope rates $s_{k}$ are linear functions of $y_{a}, y_{b}, y_{c}, y_{d}$, problem \eqref{eq:LR} constitutes a linear least squares (LS) problem, which can be solved very efficiently.

Specifically, the linear coefficients of $y = \begin{bmatrix}
    y_a \ y_b \  y_c \  y_d
\end{bmatrix}^{\mathsf T}$ for each piece can be stacked and given in matrix form:
\begin{align}
    {A}_{1} = \begin{bmatrix}
    v^{1}_{1} \ w^{1}_{1} \ 0 \ 0\\ 
    \cdots \\
    v^{i}_{1} \ w^{i}_{1} \ 0 \ 0\\
    \vdots
    \end{bmatrix}
    {A}_{2} = \begin{bmatrix}
    0 \ v^{1}_{2} \ w^{1}_{2} \ 0\\ 
    \cdots \\
    0 \ v^{i}_{2} \ w^{i}_{2} \ 0\\
    \vdots
    \end{bmatrix}
    {A}_{3} = \begin{bmatrix}
    0 \ 0 \ v^{1}_{3} \ w^{1}_{3}\\ 
    \cdots \\
    0 \ 0 \ v^{i}_{3} \ w^{i}_{3}\\
     \vdots
    \end{bmatrix} \nonumber
\end{align}
where $v^{i}_{k}, w^{i}_{k}, \forall i \in \{1,2,\cdots,N_k\}$ represent nonzero coefficients corresponding to the $i$-th sample in piece $k$, and can be explicitly given by:
\begin{subequations}
\begin{align} 
 v^{i}_{1} = \frac{x_b - I_i}{x_b - x_a}, v^{i}_{2} &= \frac{x_c - I_i}{x_c - x_b}, v^{i}_{3} = \frac{x_d - I_i}{x_d - x_c}\\  
 w^{i}_{1} = \frac{I_i - x_a}{x_b - x_a},
w^{i}_{2} &= \frac{I_i - x_b}{x_c - x_b},
w^{i}_{3} = \frac{I_i - x_c}{x_d - x_c}
\end{align}
\end{subequations}
Concatenation of coefficients and constants in \eqref{eq:LR} leads to:
\begin{align} \label{eq:LS_1}
    {A} = \begin{bmatrix}
   {A}_{1} \\ 
   {A}_{2} \\
   {A}_{3}
    \end{bmatrix}, \quad 
    b = \begin{bmatrix}
   {V}_{1} \\ 
   \vdots \\
   {V}_{N}
    \end{bmatrix} = V
\end{align}
To this end, the closed-form solution to optimization problem \eqref{eq:LR} immediately follows:
\begin{align} \label{eq:LS_2}
        {y} = ({A}^{\mathsf T} {A})^{-1} {A}^{\mathsf T} {b}
\end{align}
Overall, the proposed LS method is very efficient in handling noisy data and outliers, and can be conveniently modified to accommodate more segments as needed.

\subsection{Piecewise Quadratic Approximation}

\begin{figure}[t!]
\centering
\includegraphics[trim=0cm 0cm 0cm 0.5cm,clip=true,totalheight=0.16\textheight]{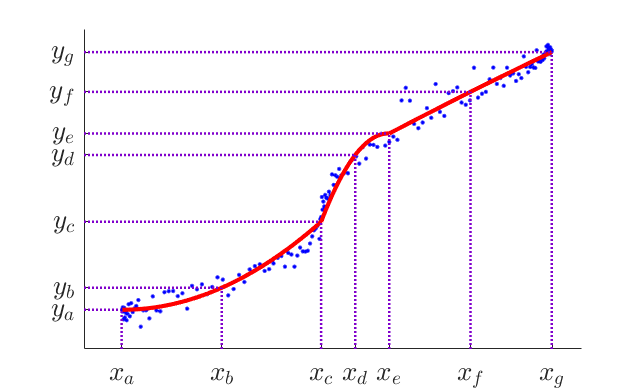}
\caption{Illustration of the piecewise quadratic approximation}
\label{fig:approximation_quad}
\vspace*{-3mm}
\end{figure}

To improve the fitting performance, one can further consider piecewise quadratic approximation (see Fig. \ref{fig:approximation_quad}). The goal then becomes finding three optimal quadratic functions:
\begin{align}
y_{k} = m_{k}x^{2} + n_{k}x + h_{k}, \quad \forall k = 1, 2, 3.
\end{align}
where $m_k, n_k, h_k$ denote quadratic, linear, and constant coefficients, respectively. Different from the linear approximation, uniquely determining a quadratic function requires at least three points. Therefore, an additional interpolation point needs to be introduced for each piece. For easier derivation of the coefficients $m, n$ later, one can pick the midpoint of each piece for interpolation. 
Hence, the problem can be formulated as:
\begin{subequations} \label{eq:QR}
\begin{align} 
\min \quad & \sum^{3}_{k = 1} \sum^{N_k}_{i = 1}  \left(m_{k}I^{2}_{i} + n_{k}I_{i} + y^{*}_{k} - m_{k}{x^{*}_{k}}^{2} - n_{k}x^{*}_{k} - V_{i}\right)^{2} \label{eq:QA_a}\\
\textrm{s.t.} \quad 
  & \underline{y}_{\zeta} \leq y_{\zeta}  \leq \overline{y}_{\zeta}, \quad \zeta = a, b, c, d, e, f, g. \label{eq:QA_c}
\end{align}
\end{subequations}
where the number of decision variables is increased to include additional interpolation points $x_b$, $x_d$ and $x_f$ (see Fig. \ref{fig:approximation_quad}).
Since the midpoint of each piece is selected as the interpolation point, the quadratic and linear coefficients $m, n$ can be conveniently derived. For example, for the first piece $k = 1$, calculating $(y_c - y_b) - (y_b - y_a)$ cancels out the linear and constant coefficients, and thus:
\begin{align}
m_{1} = \frac{y_{c}-2y_{b}+y_{a}}{x^{2}_{c}-2x^{2}_{b}+x^{2}_{a}}
\end{align}
Further, combining this equation with $y_b - y_a = m_1 (x^{2}_b - x^{2}_a) + n_1 (x_b - x_a)$ gives us the linear coefficient:
\begin{align}
n_{1} = \frac{y_{b}-y_{a} - m_{1}(x^{2}_{b} - x^{2}_{a})}{x_{b}-x_{a}}
\end{align}
Similarly, the coefficients $m, n$ for other pieces can be obtained.
Note that both quadratic and linear coefficients are linear functions of $y$, hence problem \eqref{eq:QR} is also a linear LS problem that is efficiently solvable (similar to \eqref{eq:LS_1}-\eqref{eq:LS_2}).
Of course, the quadratic model provides superior fitting performance because the linear function is just a special case of the quadratic function. However, as higher-order modeling inevitably introduces more variables, a slightly higher computational complexity will incur as a trade-off.

\begin{remark}[On parametric continuity]
Constraints on first-order parametric continuity are not enforced for the piecewise approximation because there is also no such guarantee on the V-I trajectory during the arc quenching and restriking.
\end{remark}

\section{Support Vector Machine Approach} \label{sec:svm}

The approximation algorithms introduced earlier simplify the nonlinear V-I trajectory of HIFs to piecewise functions (linear, quadratic), which makes it easier to fit into the machine learning framework.
Specifically, the approximated piecewise functions can effectively reflect the system's electrical characteristics during the arc restriking and quenching. Therefore, a collection of features (e.g., slope rates) of these approximated functions can serve as the learning inputs to identify the HIF locations. Compared with other supervised learning algorithms such as neural networks, the SVM approach delivers more explainable and interpretable solutions for HIF identification.

SVM is an efficient supervised learning algorithm used for classification and regression tasks. The algorithm seeks to find a set of hyperplanes, ${w}^\mathsf{T} {x} - b = 1$ and ${w}^\mathsf{T} {x} - b = -1$ (e.g., linear SVM) that maximize the separation margin $\frac{2}{||{w}||}$, which can be formulated as an optimization problem:
\begin{subequations}
\begin{align} 
\min \quad & ||{w}||^{2}\\
\textrm{s.t.} \quad &  y_{i}({w}^{\mathsf T}x_{i} - b) \geq 1, \: \forall i
\end{align}
\end{subequations}
Going beyond the simple two-class linear model, advances in SVM algorithms in the past few decades have enabled multiclass SVM to be easily constructed. Furthermore, by applying the well-known kernel trick \cite{scholkopf2000kernel}, it is also possible to perform nonlinear classification.

When SVM is used for the HIF identification task, the learning output $y$ is the fault location label (e.g., bus number), and we propose utilizing the features from the approximation functions as the input $x$. Specifically, for the piecewise linear approximation, the SVM input can be constructed as:
\begin{align} 
x_{\mathcal{L}} = \{s_1, s_2, s_3\}
\end{align}
which consists of slope rates for all segments in the piecewise linear function. Since quadratic functions contain more information than linear functions, SVM inputs under quadratic approximation will be lifted to a higher-dimensional space:
\begin{align} 
x_{\mathcal{Q}} = \{(m_1,n_1),(m_2,n_2),(m_3,n_3)\}
\end{align}
which include both quadratic and linear coefficients for each piece of the function. Note that for both $x_{\mathcal{L}}$ and $x_{\mathcal{Q}}$, the constant coefficient $h_{k}$ is not included. This is because for the V-I trajectory, the ``shape'' (e.g., $y', y''$) in general is more meaningful than the ``height'' (e.g., constant $h$) as the former substantially reflects the change in the system's electrical characteristics over the course of HIF.
The rationale behind using the approximation function features for the SVM algorithm is twofold. 
First, as the V-I trajectories for HIFs at different locations are generally different, their unique projections to emblematic piecewise functions are expected to be different as well.
Second, thanks to the piecewise approximation, the size of the learning input is greatly reduced to the order of $k$, which offers a more efficient and explainable learning model.

\begin{figure*}[hbt!]
\centering 
   \subfloat[]{
      \includegraphics[trim=0cm 0cm 0cm 0cm,clip=true, width=0.28\textwidth]{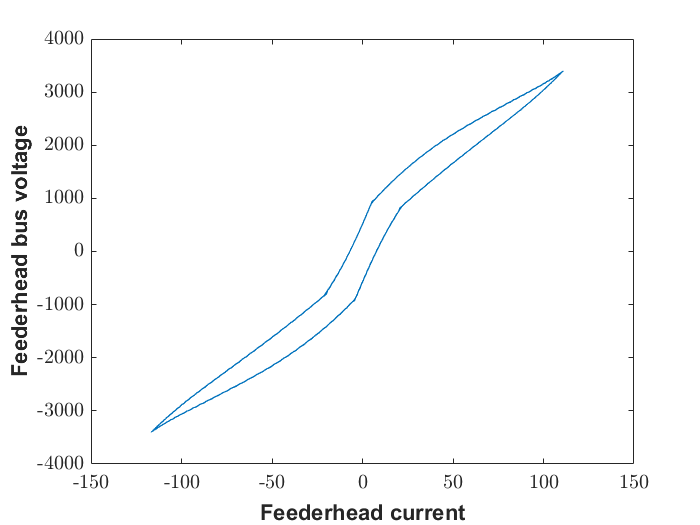} \label{fig:appr_1_1}} 
 \hspace{1.1em}
   \subfloat[]{
      \includegraphics[trim=0cm 0cm 0cm 0cm,clip=true, width=0.28\textwidth]{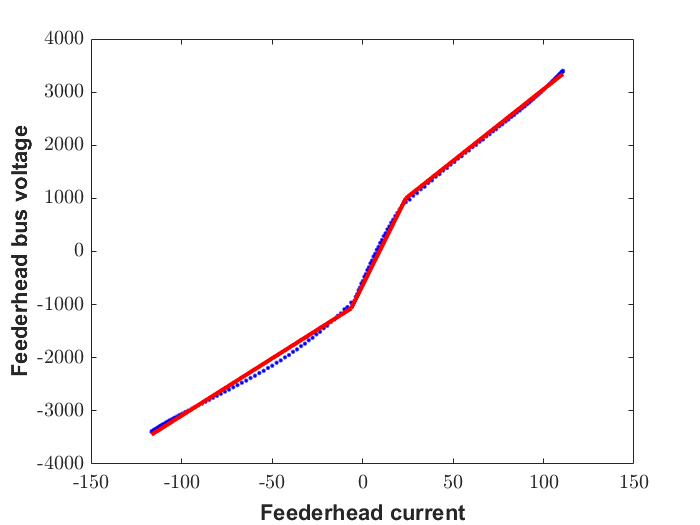}  \label{fig:appr_1_2}} 
 \hspace{1.1em}
   \subfloat[]{
      \includegraphics[trim=0cm 0cm 0cm 0cm,clip=true, width=0.28\textwidth]{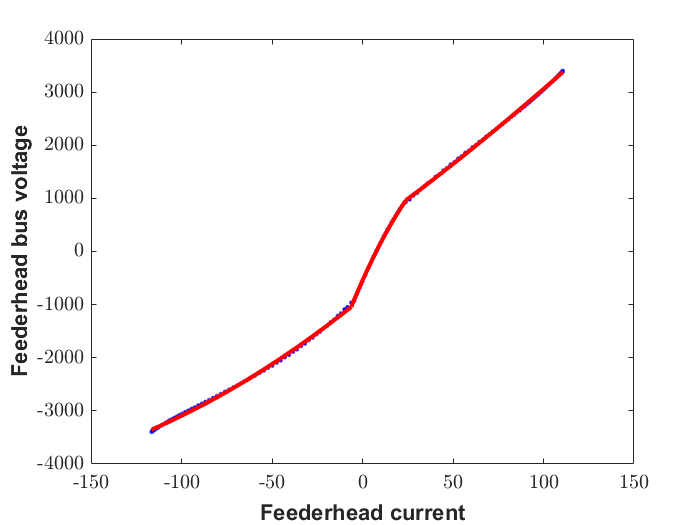} \label{fig:appr_1_3}}\\
\caption{The (a) V-I trajectory, (b) linear approximation, and (c) quadratic approximation, for an HIF at Bus 7}
\label{fig:appr_1}
\vspace*{-5mm}
\end{figure*}


\begin{figure*}[hbt!]
\centering  
   \subfloat[]{
      \includegraphics[trim=0cm 0cm 0cm 0cm,clip=true, width=0.28\textwidth]{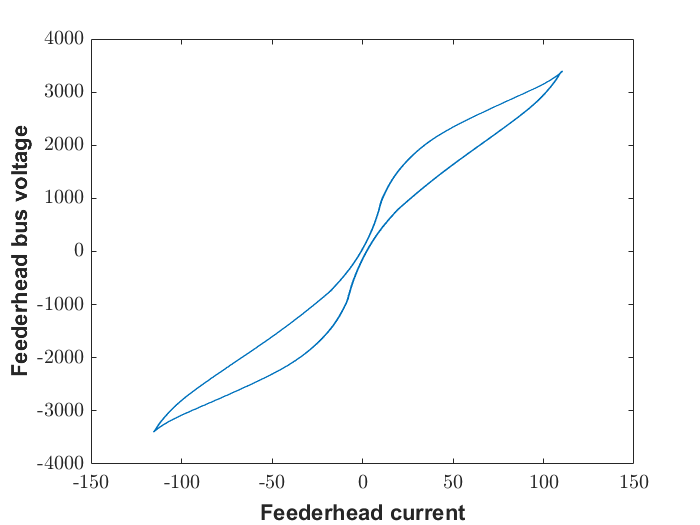}  \label{fig:appr_2_1}} 
 \hspace{1.1em}
   \subfloat[]{
      \includegraphics[trim=0cm 0cm 0cm 0cm,clip=true, width=0.28\textwidth]{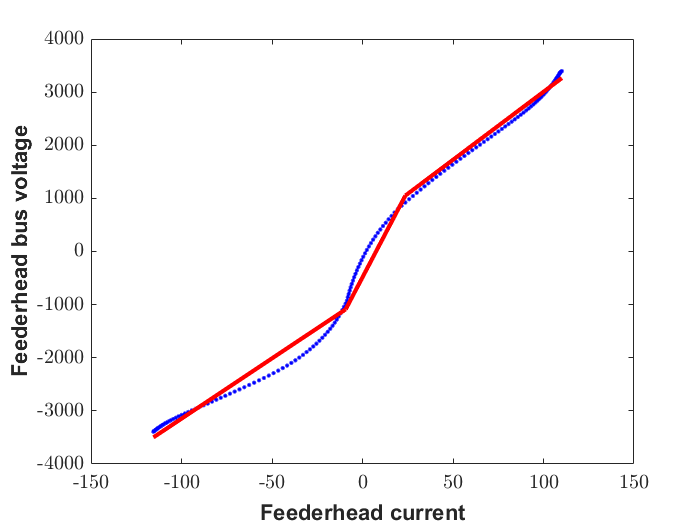} \label{fig:appr_2_2}}
 \hspace{1.1em}
   \subfloat[]{
      \includegraphics[trim=0cm 0cm 0cm 0cm,clip=true, width=0.28\textwidth]{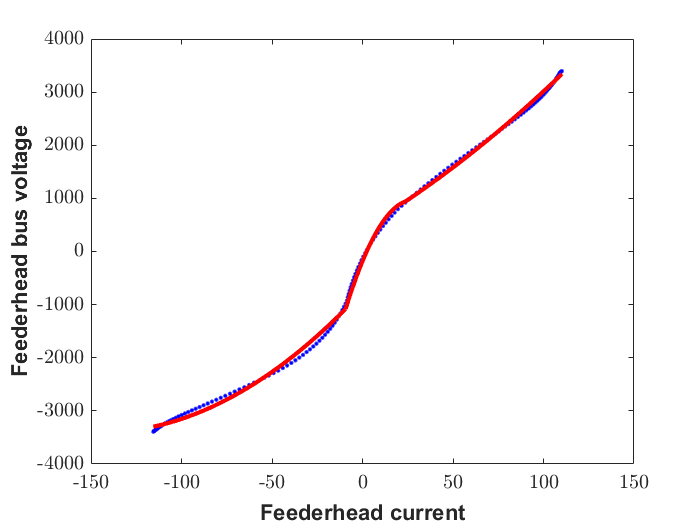} \label{fig:appr_2_3}}\\
\caption{The (a) V-I trajectory, (b) linear approximation, and (c) quadratic approximation, for an HIF at Bus 64}
\label{fig:appr_2}
\vspace*{-4mm}
\end{figure*}

\section{Numerical Results} \label{sec:NR}
This section validates the performance of the piecewise approximation and SVM algorithm for HIF identification tasks on the IEEE 123-node test feeder.
The test feeder is simulated in EMTP-ATP (ElectroMagnetic Transients Program, version Alternative Transients Program) with 14 rooftop photovoltaic (PV) units incorporated. In this model, the loading condition ($\{0.4, 1.0\}$ per unit) and PV capacity ($\{0.4, 0.6, 0.8, 1.0\}$ per unit) are varied for different steady-state scenarios. The HIF simulations span across 66 locations in total, with 63 internal nodes as well as 3 adjacent nodes outside the feeder. The generated data are sampled at 20 kHz and posted open source to the public \cite{dong2023}.
The simulations are performed on a regular laptop with Intel\textsuperscript{\textregistered} CPU @ 2.60 GHz and 16 GB of RAM. The piecewise approximation is solved using CPLEX, and SVM algorithms are implemented using the MATLAB simulator.


In this work, we primarily utilize the feederhead voltage and current measurements to validate the HIF identification algorithm. The feederhead V-I trajectories for two HIFs are shown in Fig. \ref{fig:appr_1_1} and \ref{fig:appr_2_1}. As the validated trajectory displays partial symmetry, we preprocess the data to extract only the lower half of the trajectory for the approximation and the SVM algorithms later. The piecewise linear and quadratic approximation results are demonstrated in Fig. \ref{fig:appr_1_2} (\ref{fig:appr_2_2}) and Fig. \ref{fig:appr_1_3} (\ref{fig:appr_2_3}), respectively. 
Clearly, the piecewise quadratic model provides a more accurate approximation. However, as the linear approximation offers a lower-dimensional SVM input, $x_{\mathcal{L}}$, this makes it a favorable choice for us, especially for the visualization of the SVM algorithm. 
Additionally, the slope rate for each piece of the trajectory under the linear approximation implies the ``equivalent impedance'' from the feederhead, according to Ohm's law, which can be also beneficial for our analysis.
For these reasons, we will use the results from the piecewise linear approximation to demonstrate the learning performance. Nevertheless, one can always raise the input dimension to switch to the quadratic model.

\begin{figure}[t!]
\centering
\includegraphics[trim=0cm 0cm 0cm 0cm,clip=true,totalheight=0.15\textheight]{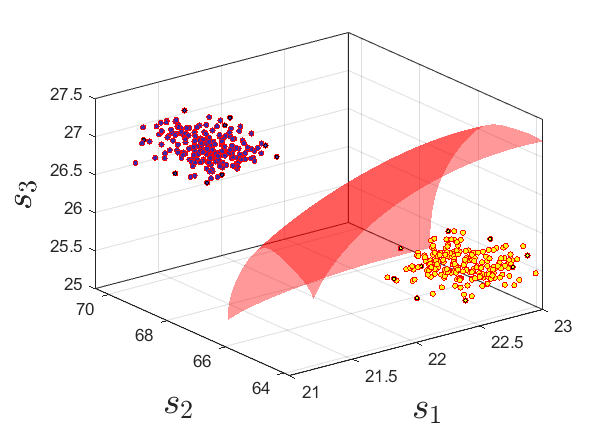}
\caption{SVM on HIFs under the three-dimensional space}
\label{fig:svm_1}
\vspace*{-4mm}
\end{figure}

\begin{figure}[t!]
\centering
\includegraphics[trim=0cm 0cm 0cm 0cm,clip=true,totalheight=0.15\textheight]{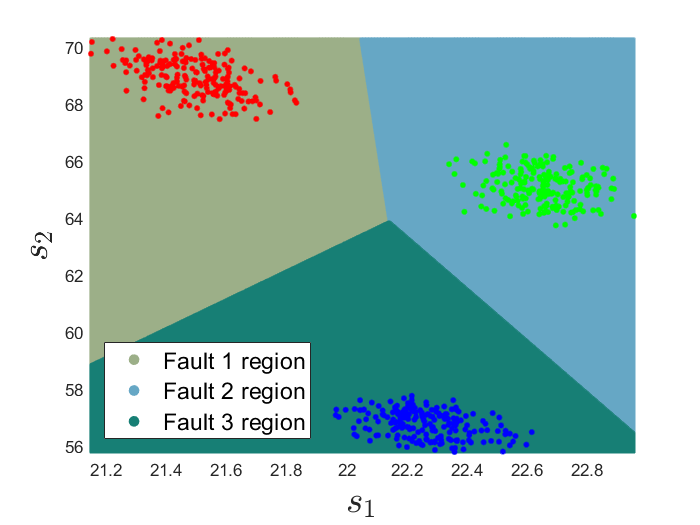}
\caption{SVM on HIFs under the two-dimensional space}
\label{fig:svm_2}
\vspace*{-4mm}
\end{figure}

The actual V-I trajectory might exhibit non-smoothness due to measurement errors, and thus we add noise proportional to the nominal measurement value to simulate this effect. 
The presence of noise in trajectories produces different approximation results and correspondingly leads to dispersed SVM inputs $x_{\mathcal{L}} = \{s_1, s_2, s_3\}$ in the three-dimensional space. The classification of two faults at buses 7 and 64 using the Gaussian kernel as an example is shown in Fig. \ref{fig:svm_1}. In fact, the data points are already linearly separable because the two HIFs are located far away from each other. The hyperplane in red distinctly classifies these two faults using the slope rates of V-I trajectories. Despite the similarity presented in the approximation results in Fig. \ref{fig:appr_1} and Fig. \ref{fig:appr_2}, they are clearly identified under three-dimensional space. 
For the SVM under multiple HIFs, we can project the input to a two-dimensional space by keeping only two slope rates ($s_1$ and $s_2$). The performance of SVM with a linear kernel on three different HIFs is demonstrated in Fig. \ref{fig:svm_2}. These faults are simulated at buses 7, 64, and 82, respectively. Because they are located in different areas of the system, even with partial inputs $\{s_1, s_2\}$, the linear SVM can efficiently identify these HIFs.

\begin{figure}[t!]
\centering
\includegraphics[trim=0cm 0cm 0cm 1cm,clip=true,totalheight=0.15\textheight]{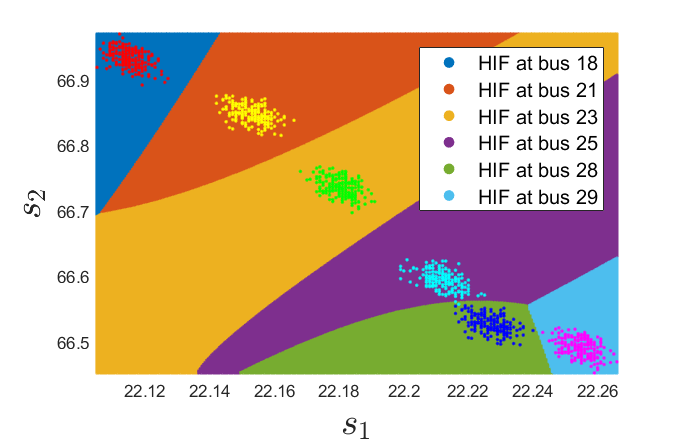}
\caption{SVM on HIFs located on the same branch}
\label{fig:svm_3}
\vspace*{-4mm}
\end{figure}

In contrast, HIFs located in the same area are generally more difficult to distinguish due to their similar network topology and electrical characteristics. To that end, we also validate the identification performance on six HIFs that are located on the same branch of the system. These faults, based on their distance to the feederhead, are at buses 18, 21, 23, 25, 28, and 29 (from nearest to farthest), respectively. The results of the SVM with a polynomial kernel under two-dimensional space are shown in Fig. \ref{fig:svm_3}. Because these faults are adjacent, the slope rates $\{s_1,s_2\}$ of their V-I trajectories are relatively close. 
The regions corresponding to distinct faults are colored differently, as denoted in the figure legend.
Interestingly, as the fault location moves farther from the feederhead, the slope rate, $s_1$, which corresponds to the arc restriking phase of the HIF, gets bigger. This is mainly because when the fault is located deeper on that branch, fewer system components will be disconnected behind that fault, and therefore the equivalent impedance, $Z = V/I$ (or slope rate $s_1$), is bigger. Then again, the results have demonstrated that compared with area-based HIF identification, bus-level localization of HIFs in small areas is possible but more challenging in general.

\section{Conclusions and Future Work}
\label{sec:con}
In this paper, we propose an explainable data-driven approach to localize HIFs efficiently. To handle the nonlinearity of the HIF, we approximate the V-I trajectory with piecewise linear (or quadratic) functions. Following that, we feed the simplified function features as inputs and utilize the SVM method for HIF identification.
Numerical tests on the IEEE 123-bus distribution system demonstrate the validity and efficiency of the proposed algorithm. Further work includes higher-order approximation and learning algorithms under system uncertainties (e.g., renewable resources, topology change).




%

\bibliography{bibliography.bib}

\bibliographystyle{IEEEtran}

\itemsep2pt

\end{document}